\documentclass[english,preprint,showpacs,preprintnumbers]{revtex4}
\usepackage[T1]{fontenc}
\usepackage[latin9]{inputenc}
\usepackage{amsmath}
\usepackage{amssymb}

\makeatletter
\@ifundefined{textcolor}{}
{%
 \definecolor{BLACK}{gray}{0}
 \definecolor{WHITE}{gray}{1}
 \definecolor{RED}{rgb}{1,0,0}
 \definecolor{GREEN}{rgb}{0,1,0}
 \definecolor{BLUE}{rgb}{0,0,1}
 \definecolor{CYAN}{cmyk}{1,0,0,0}
 \definecolor{MAGENTA}{cmyk}{0,1,0,0}
 \definecolor{YELLOW}{cmyk}{0,0,1,0}
 }

\usepackage{amsfonts}\usepackage{graphics}\usepackage{epsfig}\usepackage{dcolumn}\usepackage{bm}

\makeatother

\usepackage{babel}
\begin{document}

\title{Neutrino masses and the scalar sector of a $B-L$ extension of the
standard model.}

\author{J. C. Montero}

\email{montero@ift.unesp.br}

\selectlanguage{english}%

\affiliation{Instituto de Física Teórica--Universidade Estadual Paulista \\
 R. Dr. Bento Teobaldo Ferraz 271, Barra Funda\\
 São Paulo - SP, 01140-070, Brazil }

\author{B. L. Sánchez--Vega}

\email{brucesan@ift.unesp.br}

\selectlanguage{english}%

\affiliation{Instituto de Física Teórica--Universidade Estadual Paulista \\
 R. Dr. Bento Teobaldo Ferraz 271, Barra Funda\\
 São Paulo - SP, 01140-070, Brazil }
\begin{abstract}
We consider an electroweak model based on the gauge symmetry $SU(2)_{L}\otimes U(1)_{_{Y^{\prime}}}\otimes U(1)_{_{B-L}}$
which has right-handed neutrinos with different exotic B-L quantum
numbers. Because of this particular feature we are able to write Yukawa
terms, and right-handed neutrino mass terms, with scalar fields that
can develop vacuum expectation values belonging to different energy
scales. We make a detailed study of the scalar and the Yukawa neutrino
sectors to show that this model is compatible with the observed solar
and atmospheric neutrino mass scales and the tribimaximal mixing matrix.We
also show that there are dark matter candidates if a $Z_{2}$ symmetry
is included.
\end{abstract}

\pacs{14.60.St,\,12.60.Cn,\,12.60.Fr }

\maketitle

\section{Introduction}

\label{sec:intro} The neutrino masses and mixing which are required
for giving a consistent explanation for the solar and atmospheric
neutrino anomalies are the most firm evidence of physics beyond the
electroweak standard model (ESM). New physics can be implemented in
a variety of different scenarios. There are basically two main schemes
that are often followed: (i) new matter content is added to the model
respecting the original ESM gauge symmetry and (ii) to consider a
model with a larger gauge symmetry. Certainly both schemes can be
implemented together. In this vein, extensions of the ESM having an
extra $U(1)$ gauge symmetry factor are interesting for a variety
of reasons. They are the simplest way of extending the ESM gauge symmetry
and can be thought of as an intermediate energy scale symmetry coming
from the breaking, at a higher energy scale, of a larger gauge symmetry
describing some yet unknown physics. For instance, $U(1)$ gauge factors
are contained in grand unified theories, supersymmetric models, and
left-right models. One major feature of these models is the existence
of an extra neutral vector boson, usually denoted by $Z^{\prime}$,
whose mass is related to the energy scale of the extra $U(1)$ symmetry
spontaneously broken. It is expected to have $Z^{\prime}$ signals
at the TeV scale and its discovery is one of the goals of the LHC
and future lepton colliders. Depending on the implementation of this
kind of model, it can have a natural candidate for dark matter (DM)
and/or furnish a mechanism for leptogenesis. Through the years much
work has been done considering the features of this extra $U(1)$
gauge factor and some particular formulations of the model were made.
See, for example, Refs.~\cite{appelquist,carlson}. In particular,
when the charge of the extra $U(1)$ factor is identified with $B-L$
(baryon number minus lepton number), there is extensive literature
concerning the most different versions of the model and a large variety
of phenomenological aspects.

In this paper we consider a $B-L$ gauge model which has the particularity
of being rendered anomaly free by introducing right-handed neutrinos
with exotic $B-L$ charges. The number of right-handed neutrinos and
their $B-L$ exotic charges is fixed by the anomaly cancellation equations.
Since in this model not all of these right-handed neutrinos have the
same exotic charge, we can construct Yukawa terms with different $SU(2)_{L}$
scalar doublets. Appropriate $SU(2)_{L}$ scalar singlets are also
introduced to write the most general mass terms for the right-handed
neutrinos. We make a detailed study of the scalar potential, concerning
the mass spectra and the physical Goldstone bosons, and take advantage
of this rich scalar sector to construct a seesaw mechanism at low
energies (TeV scale) to give realistic masses to the light active
neutrinos.

The outline of this paper is as follows. In Sec. II we present the
particular $B-L$ gauge model under consideration. In Sec. III we
analyze the scalar potential of the model --the symmetries, the mass
spectra and the model compatibility with experimental constraints
--and introduce a $Z_{2}$ symmetry to allow the model to have DM
candidates. In Sec. IV we study the neutrino mass generation and show
the compatibility of the model with the observed neutrino masses and
the tribimaximal mixing. Finally, our conclusions are given in Sec.
V.

\section{The model}

\label{sec:flipped} We consider the model of Ref.~\cite{montero}
that we briefly summarize here. The model is an extension of the ESM
based on the gauge symmetry $SU(2)_{L}\otimes U(1)_{Y^{\prime}}\otimes U(1)_{B-L}$
where $B$ and $L$ are the usual baryonic and leptonic numbers, respectively,
and $Y^{\prime}$ is a new charge. The values of $Y^{\prime}$ are
chosen to obtain the ESM hypercharge $Y$ through the relation $Y=[Y^{\prime}+(B-L)]$,
after the first spontaneous symmetry breaking. In order to make the
model anomaly free we have to introduce right-handed neutrinos ($n_{R}$).
Solving the anomaly equations we find that the number of $n_{R}$
cannot be less than 3, if we restrict ourselves to integer quantum
numbers only. For the minimal number (3) these equations have two
solutions: the usual one, where all right-handed neutrinos are identical
and have $L=1$, and the exotic one, where two of them have $L=4$
and the third one has $L=-5$. The model under consideration has the
right-handed neutrinos having such exotic lepton numbers.

The fermionic content of the model is the same as the ESM plus the
right-handed neutrinos introduced above. The respective charge assignment
is shown in Table~\ref{table1}. In the framework of a gauge theory
with spontaneous symmetry breaking, we at least have to introduce
a scalar doublet, $H$, in order to give mass to the lighter massive
neutral vector boson ($Z$) and the charged fermions, as in the ESM.
However, more scalar fields are needed to give mass to the extra neutral
vector boson ($Z^{\prime}$), which is expected to be heavier than
$Z$, and to the neutrinos of the model. Respecting gauge invariance,
a general choice is to introduce two $SU(2)$ scalar doublets, $\Phi_{1,2}$,
and three $SU(2)$ neutral scalar singlets, $\phi_{1,2,3}$, with
the charge assignments shown in Table~\ref{table2}.
\begin{table}[h]
\caption{Quantum number assignment for the fermionic fields.}
\label{table1} \centering
\begin{tabular}{lcccccc}
\hline\hline
$ \quad \quad$ &$\quad \quad I_{3}\quad$ & $\quad I\quad$ & $\quad Q \quad$ & $\quad Y^{\prime } \quad$ & $\quad B-L\quad$ & $\quad Y \quad \quad$ \\
\hline
$\nu _{eL}$ & $1/2$ & $1/2$ & $0$ & $0$ & $-1$ & $-1$ \\
$e_{L}$ & $-1/2$ & $1/2$ & $-1$ & $0$ & $-1$ & $-1$ \\
$e_{R}$ & $0$ & $0$ & $-1$ & $-1$ & $-1$ & $-2$ \\
$u_{L}$ & $1/2$ & $1/2$ & $2/3$ & $0$ & $1/3$ & $1/3$ \\
$d_{L}$ & $-1/2$ & $1/2$ & $-1/3$ & $0$ & $1/3$ & $1/3$ \\
$u_{R}$ & $0$ & $0$ & $2/3$ & $1$ & $1/3$ & $4/3$ \\
$d_{R}$ & $0$ & $0$ & $-1/3$ & $-1$ & $1/3$ & $-2/3$ \\
$n_{1R}$ & $0$ & $0$ & $0$ & $4$ & $-4$ & $0$ \\
$n_{2R}$ & $0$ & $0$ & $0$ & $4$ & $-4$ & $0$ \\
$n_{3R}$ & $0$ & $0$ & $0$ & $-5$ & $5$ & $0$ \\ \hline\hline
\end{tabular}%
\end{table}
With these fields, and omitting summation symbols, the most general
Yukawa Lagrangian respecting the gauge invariance is given by
\begin{eqnarray}
-\mathcal{L}_{\text{Y}} & = & Y_{i}^{(l)}\overline{L}_{Li}e_{Ri}H+Y_{ij}^{(d)}\overline{Q}_{Li}d_{Rj}H+Y_{ij}^{(u)}\overline{Q}_{Li}u_{Rj}\widetilde{H}+\mathcal{D}_{im}\overline{L}_{Li}n_{Rm}\Phi_{1}+\mathcal{D}_{i3}\overline{L}_{Li}n_{R3}\Phi_{2}\notag\\
 &  & +\mathcal{M}_{mn}\overline{(n_{Rm})^{c}}n_{Rn}\phi_{1}+\mathcal{M}_{33}\overline{(n_{R3})^{c}}n_{R3}\phi_{2}+\mathcal{M}_{m3}\overline{(n_{Rm})^{c}}n_{R3}\phi_{3}+\textrm{H.c.},\label{lyukawa}
\end{eqnarray}
where $i,j=1,2,3$ are lepton family numbers and represent $e,\,\mu$
and $\tau$, respectively, $m,n=1,2$, and $\widetilde{H}=i\tau_{2}H^{\ast}$.
The corresponding scalar potential is
\begin{eqnarray}
V_{B-L} & = & -\mu_{H}^{2}H^{\dagger}H+\lambda_{H}(H^{\dagger}H)^{2}-\mu_{11}^{2}\Phi_{1}^{\dagger}\Phi_{1}+\lambda_{11}\left\vert \Phi_{1}^{\dagger}\Phi_{1}\right\vert ^{2}-\mu_{22}^{2}\Phi_{2}^{\dagger}\Phi_{2}+\lambda_{22}\left\vert \Phi_{2}^{\dagger}\Phi_{2}\right\vert ^{2}\notag\\
 &  & -\mu_{s\alpha}^{2}\left\vert \phi_{\alpha}\right\vert ^{2}+\lambda_{s\alpha}\left\vert \phi_{\alpha}^{\ast}\phi_{\alpha}\right\vert ^{2}+\lambda_{12}\left\vert \Phi_{1}\right\vert ^{2}\left\vert \Phi_{2}\right\vert ^{2}+\lambda_{12}^{\prime}(\Phi_{1}^{\dagger}\Phi_{2})(\Phi_{2}^{\dagger}\Phi_{1})+\Lambda_{H\gamma}\left\vert H\right\vert ^{2}\left\vert \Phi_{\gamma}\right\vert ^{2}\notag\\
 &  & +\Lambda_{H\gamma}^{\prime}(H^{\dagger}\Phi_{\gamma})(\Phi_{\gamma}^{\dagger}H)+\Lambda_{Hs\alpha}\left\vert H\right\vert ^{2}\left\vert \phi_{\alpha}\right\vert ^{2}+\Lambda_{\gamma\alpha}^{\prime}\left\vert \Phi_{\gamma}\right\vert ^{2}\left\vert \phi_{\alpha}\right\vert ^{2}+\left[\Phi_{1}^{\dagger}\Phi_{2}(\beta_{13}\phi_{1}\phi_{3}^{\ast}+\beta_{23}\phi_{2}^{\ast}\phi_{3})\right.\notag\\
 &  & +\left.\beta_{123}\phi_{1}\phi_{2}(\phi_{3}^{\ast})^{2}+\textrm{H.c.}\right]+\Delta_{\alpha\beta}(\phi_{\alpha}^{\ast}\phi_{\alpha})(\phi_{\beta}^{\ast}\phi_{\beta}),\label{potential1}
\end{eqnarray}
where $\gamma=1,2$; $\alpha,\beta=1,2,3$; and $\alpha<\beta$ in
the last term. In $\mathcal{L}_{\text{Y}}$, the motivation for introducing
such scalar fields is to write the most general neutrino mass terms.
Because of the fact that not all right-handed neutrinos have the same
$Y^{\prime}$ and $(B-L)$ charges, the neutrino mass matrix will
have entries proportional to vacuum expectation values (VEVs) which
can, in principle, belong to different energy scales. The scalar potential
is a consequence of the fields we have previously introduced, and
the terms in Eq.~(\ref{potential1}) are only dictated by gauge invariance.
Now, we have to observe that when we write terms based on general
grounds, although correct, we may have introduced more symmetries
than we need. Hence, we must do a detailed study of the scalar potential,
and know the scalar mass spectra in order to avoid inconsistencies
with the present phenomenology.
\begin{table}[h]
\caption{Quantum number assignment for the scalar fields.}
\label{table2} \centering
\begin{tabular}{lcccccc}
\hline\hline
$\quad \quad$ & $\quad I_{3}\quad$ & $\quad I\quad$ & $\quad Q\quad$ & $\quad Y^{\prime}\quad$ & $\quad B-L\quad$ & $\quad Y\quad$\\
\hline
$H^{0,+}$ & $\mp1/2$ & $1/2$ & $0,1$ & $1$ & $0$ & $1$ \\
$\Phi_{1}^{0,-}$ & $\pm1/2$ & $1/2$ & $0,-1$ & $-4$ & $+3$ & $-1$\\
$\Phi_{2}^{0,-}$ & $\pm1/2$ & $1/2$ & $0,-1$ & $5$ & $-6$ & $-1$ \\
$\phi_{1}$ & $0$ & $0$ & $0$ & $-8$ & $+8$ & $0$ \\
$\phi_{2}$ & $0$ & $0$ & $0$ & $10$ & $-10$ & $0$ \\
$\phi_{3}$ & $0$ & $0$ & $0$ & $1$ & $-1$ & $0$ \\ \hline\hline
\end{tabular}%
\end{table}
The scalar doublets of the model, $H$ and $\Phi_{1,2}$, contribute
to the $Z$ boson mass, so their vacuum expectation values are bounded
by the electroweak energy scale. Hence, the largest energy scale of
the model comes from the $SU(2)$ scalar singlets $\phi_{1,2,3}$.
In this way, the pattern of the spontaneous symmetry breaking is
\begin{eqnarray}
 &  & SU(2)_{L}\otimes U(1)_{Y^{\prime}}\otimes U(1)_{B-L}\ \ \underrightarrow{\phantom{x}\left\langle \phi_{1,2,3}\right\rangle \phantom{x}}\notag\\
 &  & \qquad\qquad SU(2)_{L}\otimes U(1)_{Y}\quad\ \underrightarrow{\phantom{x}\left\langle H,\Phi_{1,2}\right\rangle \phantom{x}}\quad U(1)_{em}.
\end{eqnarray}

\section{The scalar potential analysis}

Now, we focus on the analysis of the $V_{B-L}$ scalar potential given
in Eq.~(\ref{potential1}) when all neutral scalar fields develop
nonvanishing VEVs, with the usual shifting $\varphi^{0}=\frac{1}{\sqrt{2}}(V_{\varphi}+\text{Re}\varphi+i\text{Im}\varphi)$.
By using standard procedures we are able to find the constraint equations
coming from the linear terms in the scalar potential after the symmetry
breaking. See the appendix. In the same way, we can construct the
mass-squared matrices for the charged, real, and imaginary scalar
fields. We start looking at the mass-squared matrix for the charged
fields. It is a complete $3\times3$ symmetric matrix in the basis
($H^{+},\Phi_{1}^{+},\Phi_{2}^{+}$) that can be easily diagonalized
and, after taking into account the constraint equations, gives the
following mass spectrum: two charged Goldstone bosons
\begin{equation}
G_{W}^{\pm}=\frac{1}{\sqrt{1+\frac{V_{H}^{2}}{V_{\Phi_{2}}^{2}}+\frac{V_{\Phi_{1}}^{2}}{V_{\Phi_{2}}^{2}}}}\left(-\frac{V_{H}}{V_{\Phi_{2}}}\, H^{\pm}+\frac{V_{\Phi_{1}}}{V_{\Phi_{2}}}\,\Phi_{1}^{\pm}+\Phi_{2}^{\pm}\right),\label{goldstones_cargados}
\end{equation}
and two massive states whose expressions we are not showing by shortness.
The fields $G_{W}^{\pm}$ will be absorbed to form the longitudinal
components of the charged massive vector bosons $W^{\pm}$. The other
two physical states remain in the spectrum and are a prediction of
the model. Later in the paper we approach numerically all the mass
spectra in some different situations.

In the neutral imaginary scalar sector we have a $6\times6$ mass-squared
matrix that, after the diagonalization procedure, shows two massive
scalar and four massless fields. Two of them will become the longitudinal
components of the $Z$ and the $Z^{\prime}$ neutral vector bosons.
The other two massless states remain in the physical spectrum. We
show the two physical Goldstone bosons only in the limit where $V_{\phi_{1,2,3}}\gg V_{H},V_{\Phi_{1,2}}$
and they are given by
\begin{equation}
G_{F_{1}}^{0}\approx\sqrt{\frac{V_{\Phi_{1}}^{2}+V_{\Phi_{2}}^{2}}{V_{H}^{2}+V_{\Phi_{1}}^{2}+V_{\Phi_{2}}^{2}}}\left(\text{Im}H^{0}+\frac{V_{H}V_{\Phi_{1}}}{V_{\Phi_{1}}^{2}+V_{\Phi_{2}}^{2}}\text{Im}\Phi_{1}^{0}+\frac{V_{H}V_{\Phi_{2}}}{V_{\Phi_{1}}^{2}+V_{\Phi_{2}}^{2}}\text{Im}\Phi_{2}^{0}\right),\label{goldstonesfisicosa}
\end{equation}

\begin{eqnarray}
G_{F_{2}}^{0} & \approx & \frac{1}{\sqrt{110}}\left(7\,\text{Im}\phi_{1}+5\,\text{Im}\phi_{2}+6\,\text{Im}\phi_{3}\right).\label{goldstones_fisicosb}
\end{eqnarray}
From the expressions above we see that $G_{F_{1,2}}^{0}$ are mainly
doublet and singlet, respectively. This fact will be analyzed later
on.

For the neutral real scalar sector we have a symmetric $6\times6$
mass-squared matrix with a nonvanishing determinant. Hence the spectrum
will not contain massless states. We will not show analytical expressions
here but we will give numerical values below.

The number of Goldstone bosons we have found by doing explicit calculations
can be easily understood by studying the global symmetries of the
scalar potential before and after the spontaneous symmetry breaking
(SSB). Before the SSB, the global symmetries of the scalar potential
are (a) $SU(2)$ acting on $H$ and $\Phi_{1,2}$ doublets, (b) $U(1)$
acting on $H$ with charge $+1$, (c) $U(1)$ acting on $\Phi_{1,2}$
with charge $+1$, and (d) two independent $U(1)_{\beta,\gamma}$
transformations acting on the fields $\Phi_{1},\Phi_{2},\phi_{1},\phi_{2},\phi_{3}$
with charges $(\frac{1}{2},-\frac{1}{2},1,-1,0)$ and $(-\frac{1}{2},\frac{1}{2},0,+2,1)$,
respectively. After the SSB the global symmetries of the scalar potential
are reduced to a single $U(1)_{\alpha}$ acting on the charged components
of the doublets, $H^{\pm}$ and $\Phi_{1,2}^{\pm}$ with charges $(\pm1,\pm1)$.
Following the Goldstone theorem, the number of Goldstone bosons is
equal to the number of broken symmetry generators. In this case the
original symmetry has $(3+4\times1)=7$ generators and the remaining
symmetry has $1$. Then we must have 6 Goldstone bosons, which is
exactly the number we have found just above: two charged and four
neutral imaginary fields.

Notice that the scalar potential given in Eq.~(\ref{potential1})
corrects the one given in Eq.~(16) of Ref.~\cite{montero} in which
the terms proportional to $\Lambda_{H\gamma}^{\prime}$ are missing.
The lack of these terms alters the global symmetries under which the
scalar potential is invariant and, consequently, the number of Goldstone
bosons in the spectra. In that case the symmetries before the SSB
are (a) $O(4)$ acting on the four components of $H$, (b) $SU(2)$
acting on $\Phi_{1,2}$, (c) $U(1)$ acting on $\Phi_{1,2}$, (d)
the two $U(1)_{\beta,\gamma}$ defined above. After the SSB the remaining
symmetries are (i) $O(3)$ acting on the components $(\text{Im}H^{0},\,\text{Re}H^{+},\,\text{Im}H^{+})$
and (ii) $U(1)$ acting on $\Phi_{1,2}^{\pm}$ with charge $\pm1$.
Therefore, we are left with $(6+3+3\times1)-(3+1)=12-4=8$ Goldstone
bosons. The same result is obtained by doing explicit calculations.
From the mass-squared matrices we find that the number of Goldstone
bosons is the expected one and also that in the charged sector we
are left with four massless states given by
\begin{equation}
G_{W}^{\pm}=H^{\mp},
\end{equation}
\begin{eqnarray}
G_{C}^{\pm} & = & \frac{1}{\sqrt{V_{\text{\ensuremath{\Phi}}_{1}}^{2}+V_{\text{\ensuremath{\Phi}}_{2}}^{2}}}\left(V_{\text{\ensuremath{\Phi}}_{1}}\text{\ensuremath{\Phi}}_{1}^{\pm}+V_{\text{\ensuremath{\Phi}}_{2}}\text{\ensuremath{\Phi}}_{2}^{\pm}\right).\label{goldstones_cargados2}
\end{eqnarray}
 Hence, this result is in conflict with the present phenomenology
since there are two extra charged massless scalars in the spectrum.

Now, we return to our present analysis. Since our analysis of the
scalar potential shows the existence of two physical Goldstone bosons,
it is time to care about the safety of the model. Before that, some
remarks about the VEVs of the model are due. The $V_{\phi_{1,2,3}}$
are the largest energy scale of the model. The main contribution to
the $Z$ boson square mass comes from the doublets so that $(V_{H}^{2}+V_{\Phi_{1}}^{2}+V_{\Phi_{2}}^{2})=V_{\text{ESM}}^{2}=(246)^{2}\,\,\text{GeV}^{2}$.
The doublet $H$ is the one that couples to quarks and to charged
leptons via Yukawa interactions, and hence, $V_{H}$ must be close
to $V_{\text{ESM}}$ to give the correct tree level mass to the quark
top, as the ESM do, for an ${\mathcal{O}}(1)$ top Yukawa coupling.
We then conclude that $V_{\Phi_{1}}^{2}+V_{\Phi_{2}}^{2}\ll V_{H}^{2}$.

The major challenge to models with physical Goldstone bosons, also
called Majorons ($J$), comes from the energy loss in stars through
the processes $\gamma+e^{-}\rightarrow e^{-}+J$. This process is
used to put limits on the $\overline{e}eJ$ coupling, and it is found
that it has to be $g_{eeJ}\leq10^{-10}$ for the Sun, and $g_{eeJ}\leq10^{-12}$
for the red-giant stars~\cite{mohapatrabook}. However, the dynamics
of the red giants has not the same level of confidence as that of
the Sun, and this fact considerably weakens the second constraint.

The physical Goldstone $G_{F_{2}}^{0}$ has components only in the
$SU(2)$ singlets $\phi_{1,2,3}$, which couple only to right-handed
neutrinos. Therefore, it is safe since there is no tree level contribution
to the energy loss process. The case for $G_{F_{1}}^{0}$ is not that
simple. $G_{F_{1}}^{0}$ has a component in the ESM-like doublet $H$,
and it contributes to $\overline{e}eJ$ through $\text{Im}H^{0}$.
The components in $\Phi_{1,2}$, which couple only to neutrinos at
the tree level, pose no problem. Since in this case symmetry eigenstates
and mass eigenstates are connected by orthogonal matrices, from Eq.~(\ref{goldstones_fisicosb})
we find
\begin{equation}
\text{Im}H^{0}\approx\sqrt{\frac{V_{\Phi_{1}}^{2}+V_{\Phi_{2}}^{2}}{V_{H}^{2}+V_{\Phi_{1}}^{2}+V_{\Phi_{2}}^{2}}}G_{F_{1}}^{0}+...,
\end{equation}
and hence,
\begin{equation}
g_{eeJ}\approx\frac{Y_{e}}{\sqrt{2}}\sqrt{\frac{V_{\Phi_{1}}^{2}+V_{\Phi_{2}}^{2}}{V_{H}^{2}+V_{\Phi_{1}}^{2}+V_{\Phi_{2}}^{2}}}=\left(\frac{\sqrt{2}\, m_{e}}{V_{H}}\right)\left(\frac{V_{\Phi}}{\sqrt{2}V_{H}}\right)\approx2\times10^{-6}\frac{V_{\Phi}}{V_{H}}\leq10^{-12}-10^{-10},\label{vlimite}
\end{equation}
where $Y_{e}$ is the electron Yukawa coupling to the doublet $H$,
we have defined $V_{\Phi_{1}}^{2}+V_{\Phi_{2}}^{2}\equiv V_{\Phi}^{2}$,
and we have used the shift $H^{0}\rightarrow\frac{1}{\sqrt{2}}(V_{H}+\text{Re}H^{0}+i\,\text{Im}H^{0})$.
From the equation above we conclude that the VEVs of the $SU(2)$
doublets $\Phi_{1,2}$ must be less than $12$ MeV to satisfy the
Sun constraint or less than $120$ KeV to satisfy rigorously the red-giant
constraint. Let us adopt for practical purposes an intermediate scale:
$V_{\Phi}=1$ MeV.

Once we have established the energy scale of the VEVs of the model,
and verified its safety up to now, we can make an exemplary study
of the full scalar mass spectra. We will do it numerically since the
excessive length of the analytical expressions make them useless.

In order to compute the masses we consider a set of parameters: the
dimensional ones $V_{H}\!=\!246,V_{\Phi_{1,2}}\!=\!0.001,V_{\phi_{1,2,3}}\!=\!1000,\ \text{in GeV, and }\lambda_{H}\!=0.2,\,\lambda_{11,22}\!=\!\Lambda_{13,21,22,H1,H2}^{\prime}\!=\!\lambda_{s\alpha}\!=\!1,\,\Lambda_{H1,H2}\!=\!\Lambda_{Hs\alpha}\!=\!\Lambda_{11,12,23}^{\prime},\,\Delta_{12,13,23}\!=0.1,\,\,\beta_{123}\!=-0.8$,
which are dimensionless, for $\alpha\!=\!1,2,3$. Note that the values
for the $\mu$ parameters are found by solving the constraint equations
for the scalar potential: $\mu_{H}^{2}\!=\!(402)^{2},\,\mu_{11}^{2}\!=\!-(630)^{2},\,\mu_{22}^{2}\!=\!(230)^{2},\,\mu_{s1}^{2}\!=\!\mu_{s2}^{2}\!=\!(838)^{2},\,\mu_{s3}^{2}\!=\!(550)^{2},\text{in GeV}^{2}$.
We also use $g_{Y^{\prime}}=g_{B-L}=0.4885$, and $g=0.6298$, where
$g_{Y^{\prime}}$, $g_{B-L}$, and $g$ are the coupling constants
of the $U(1)_{Y^{\prime}}$, $U(1)_{B-L}$, and $SU(2)_{L}$ gauge
factors, respectively, which are related to the electric charge through
$1/e^{2}=1/g_{Y^{\prime}}^{2}+1/g_{B-L}^{2}+1/g^{2}$~\cite{montero}.

The charged scalar sector gives the masses $m_{C_{j}}=(1424.9,\,173.9,\,0)$,
also in GeV, where the massless complex field is responsible for the
longitudinal components of the charged vector bosons $W^{+}\,\text{and}\,\, W^{-}$.

In the imaginary neutral scalar sector we have a $6\times6$ square
mass matrix and after diagonalization we find, in GeV, $m_{I_{i}}=(1549.2,\,1414.13,\,0,\,0,\,0,\,0)$.
Notice the correct number of the massless fields: two are absorbed
to form the longitudinal component of the neutral vector bosons of
the model, $Z$ and $Z^{\prime}$, and the other two are the physical
Goldstone bosons $G_{F_{1,2}}^{0}$, as discussed above.

In the real neutral scalar sector, in the same way, we find $m_{R_{i}}=(1743.9,\,1643.2,\,1414.2,\,1029.8,\,150.0,\,0.0014)$,
in GeV. We have found a very light scalar of about $1.4$ MeV which
poses a new challenge to the model: the $Z$ invisible decay width.
The presence of such a light scalar field, say $R$, and the $G_{F_{1}}^{0}\equiv J$
zero mass state, allows the decay $Z\rightarrow RJ\rightarrow JJJ$,
which will contribute to the $Z$ invisible decay width as half of
the decay $Z\rightarrow\overline{\nu}\nu$, for a single flavor family~\cite{conchanir}.
According to the experimental data there is no room for such an extra
contribution~\cite{pdg}.

The light scalar we found above is not the result of a particular
choice of the input parameters, as it could be thought at the first
moment. Let us provide a qualitative but convincing argument. As was
observed in Ref.~\cite{conchanir}, the reason is as follows. We
have mentioned above that, before the SSB, the scalar potential has
a $U(1)$ global symmetry acting on each of the $\Phi_{1,2}$ doublets,
say, $\Phi$. This means that we can rotate freely in the $\text{Re}\Phi^{0}$--$\text{Im}\Phi^{0}$
plane, so that as long as this $U(1)$ symmetry holds, the fields
$\text{Re}\Phi^{0}$ and $\text{Im}\Phi^{0}$ are mass degenerate.
However, this symmetry is broken when the real neutral component acquires
a nonvanishing VEV and, hence, the fields are no longer mass degenerate.
The square mass difference must be, then, of the order of the square
of the energy scale responsible for breaking the symmetry, i.e., $m_{\text{Re}\Phi^{0}}^{2}-m_{\text{Im}\Phi^{0}}^{2}=\mathcal{O}(V_{\Phi}^{2})$.
When $\text{Im}\Phi^{0}$ is a Goldstone, $m_{\text{Im}\Phi^{0}}^{2}=0$,
we are left with $m_{\text{Re}\Phi^{0}}^{2}=\mathcal{O}(V_{\Phi}^{2})$,
which, in our case, it is a very light scalar since $V_{\Phi}$ must
be of the order of $1$ MeV, in order to be consistent with the star
energy loss data. Then, we must find a way to reconcile the present
model with the experimental constraints.

Some attempts can be made to remove such inconsistency. Since the
origin of the problem is in the breaking of the $U(1)$ symmetry acting
on the doublets $\Phi_{1,2}$, let us consider the situation where
$V_{\Phi_{2}}=0$, and all other VEVs are different from zero. In
this case we find the same number of neutral Goldstone bosons (4):
two would be Goldstone bosons and two physical ones $G_{F_{1,2}}^{0}$.
$G_{F_{2}}^{0}$ is given by the same expression as in Eq.~(\ref{goldstones_fisicosb}),
and
\begin{equation}
G_{F_{1}}^{0}\approx\frac{1}{\sqrt{V_{H}^{2}+V_{\Phi_{1}}^{2}}}\left(V_{\Phi_{1}}\text{Im}H^{0}+V_{H}\text{Im}\Phi_{1}^{0}\right).\label{goldstones_fisicos1}
\end{equation}
We also find that for the same input parameters, but now providing
an input value for $\mu_{22}^{2}=(230)^{2}\,\,\text{GeV}^{2}$, the
mass spectra are practically not affected and we still have a light
real scalar whose mass is about $1$ MeV$\sim\mathcal{O}(V_{\Phi_{1}})$.
We get the same conclusion if we consider $V_{\Phi_{1}}=0\text{ and }V_{\Phi_{2}}\neq0$.
We only have to do the replacement $\Phi_{1}\longleftrightarrow\Phi_{2}$
in the above results.

As the problem persists, let us now consider the case where $V_{\Phi_{1}}=0$
and $V_{\Phi_{2}}=0$. In this case, the number of Goldstone bosons
is reduced to $3$. There is only one physical Goldstone, the $G_{F_{2}}^{0}$
given in Eq.~(\ref{goldstones_fisicosb}), which is safe, as discussed
above. The mass spectra are now considerably affected. For the same
input parameters as above, and with $\mu_{11}^{2}=-(800)^{2},\mu_{22}^{2}=(230)^{2}$,
in $\text{GeV}^{2}$, the spectra, with all the masses in GeV, are
the following. For the charged scalars we have $m_{C_{j}}=(1469.4,\,380.1,\,0)$.
For the imaginary scalars we find $m_{I_{i}}=(1549.2,\,1459.1,\,337.9,\,0,\,0,\,0)$,
and for the real scalars $m_{R_{i}}=(1743.9,\,1643.2,\,1459.1,\,1029.8,\,337.8,\,150.0)$.
As we can see, there is no a light real scalar anymore. The lighter
real scalar is heavier than the $Z$ vector boson, so that the problematic
decay $Z\rightarrow RJ$ is kinetically forbidden. Then, we have succeed
in making the model safe. However, this solution is not satisfactory
since the choice we have made for the doublet VEVs ($V_{\Phi_{1,2}}=0$)
does not allow the light neutrinos to get mass. It is easy to see
that in this case there is a remaining $U(1)$ quantum symmetry, say,
$U(1)_{\zeta}$, protecting the neutrino mass generation at any level.
A possible $\zeta$-charge assignment is: $\zeta(\nu_{eL},e_{L},e_{R},\Phi_{1,2})=-1$,
$\zeta(u_{L},d_{L},u_{R},d_{R})=1/3$, and $\zeta(n_{(1,2,3)R},\phi_{1,2,3})=0$.
In order to make the model compatible with the experimental data and,
hence, with massive neutrinos, we have to look for a new kind of solution
since the symmetry breaking pattern above is not realistic.

Before continuing the search for a satisfactory solution, let us observe
that before the SSB the model has a $Z_{2}$ exact symmetry with the
transformation rules $Z_{2}(n_{R3})=-n_{R3},\, Z_{2}(\Phi_{2})=-\Phi_{2},\, Z_{2}(\phi_{3})=-\phi_{3}$,
and all the other fields being even under $Z_{2}$. It is interesting
to preserve this symmetry after the SSB if we are looking for DM candidates.
This is true when $V_{\Phi_{2}}=V_{\phi_{3}}=0$. In this case the
$Z_{2}$ symmetry is not spontaneously broken, and a mechanism similar
to that of Ref.~\cite{ma73} can be implemented. The number of Goldstone
bosons is 4, and the physical ones are given by the following: $G_{F_{1}}^{0}$
is given by the same expression in Eq.~(\ref{goldstones_fisicos1}),
and
\begin{equation}
G_{F_{2}}^{0}=\frac{1}{\sqrt{16V_{\phi_{1}}^{2}+25V_{\phi_{2}}^{2}}}\left(5V_{\phi_{2}}\text{Im}\phi_{1}+4V_{\phi_{1}}\text{Im}\phi_{2}\right).\label{goldstones_fisicos2}
\end{equation}
 However, as we already know, there is a very light real scalar that,
together with $G_{F_{1}}^{0}$, has severe implications on the $Z$
invisible decay width. We will come to this $Z_{2}$ picture later
on.

\subsection{The solution}

With the aim of constructing a consistent model, let us introduce
a new $SU(2)$ neutral scalar singlet $\phi_{X}$ with the quantum
numbers $Y^{\prime}=-(B-L)=3$. The Yukawa Lagrangian remains as in
Eq.~(\ref{lyukawa}), but to the scalar potential in Eq.~(\ref{potential1}),
besides extending the range of the indices to $\alpha,\beta=1,2,3,X$,
we have to add the following non-Hermitian terms,
\begin{equation}
V_{B-L}^{X}=-i\kappa_{H1X}\Phi_{1}^{T}\tau_{2}H\phi_{X}-i\kappa_{H2X}(\Phi_{2}^{T}\tau_{2}H)(\phi_{X}^{\ast})^{2}+\beta_{X}(\phi_{X}^{\ast}\phi_{1})(\phi_{2}\phi_{3})+\beta_{3X}(\phi_{X}^{\ast}\phi_{3}^{3})+\textrm{H.c.},\label{potentialx}
\end{equation}
in order to account for all the gauge invariant terms after the introduction
of $\phi_{X}$. The terms above reduce the number of global symmetries
of the scalar potential, so that changes in the scalar spectra are
expected.

Before the SSB, the global symmetries of the total scalar potential
are (a) $SU(2)$ acting on $H$ and $\Phi_{1,2}$ doublets, (b) $U(1)_{\alpha}$
acting on $H\,\text{and}\,\Phi_{1,2}$, and (c) $U(1)_{\beta}$ acting
on the fields $H,\Phi_{1},\Phi_{2},\phi_{1},\phi_{2},\phi_{3},\phi_{X}$
with charges $(\frac{3}{8},0,-\frac{9}{8},1,-\frac{5}{4},-\frac{1}{8},-\frac{3}{8})$,
respectively. After the SSB the global symmetries of the scalar potential
are reduced to a single $U(1)$ acting on the charged components of
the doublets, $H^{\pm}$ and $\Phi_{1,2}^{\pm}$, with charges $(\pm1,\pm1)$.
The total number of Goldstone bosons will be given by the number of
broken generators, i.e., $5-1=4$, which is the number of massless
fields needed to form the longitudinal components of the charged ($W^{+},W^{-}$)
and neutral vector bosons ($Z,Z^{\prime}$). In this case, there are
no physical Goldstone bosons at all. It means that the inclusion of
the $SU(2)$ scalar singlet $\phi_{X}$ has removed all physical massless
states from the spectrum, and we have succeeded in finding a solution
for the safety of the model.

Now, numerical applications require expanding the input parameters
set to account for the new ones related to $\phi_{X}$. We then choose
$V_{\phi_{X}}=1000,\,\text{and}\,\kappa_{H1X}=0.01$ in GeV, and the
dimensionless $\lambda_{sX}=\Lambda_{1X}^{\prime}=1,\,\Lambda_{HsX}=\Lambda_{2X}^{\prime}=\Delta_{1X}=\Delta_{2X}=\Delta_{3X}=\beta_{3X}=0.1,\,\beta_{X}=-0.6,\text{and}\,\kappa_{H2X}=0.001$.
As before the $\mu$ parameters are found by solving the constraint
equations given in the appendix. With the above parameter set, plus
the one we have used previously, we find $m_{C_{j}}=(11\,137.3,\,1661.7,\,0)$
for the charged scalar sector, $m_{I_{i}}=(11\,135.9,\,1652.6,\,1467.0,\,973.6,\,0.002,\,0,\,0)$
for the neutral imaginary sector, and $m_{R_{i}}=(11\,135.9,\,1927.6,\,1816.6,\,1652.7,\,1508.8,\,900.5,\,146.2)$
for the real scalar sector, in GeV. Notice that we have now a very
light pseudoscalar, which has components mainly in the $SU(2)$ singlet
fields $\phi_{1,2,3,X}$. For instance, its component in $\text{Im}H^{0}$
is $7.3\times10^{-12}$, which implies $g_{eeJ}\approx10^{-18}$.
Hence, it is compatible with the astrophysical constraint, and poses
no problem to the $Z$ invisible decay width, since all the real scalar
fields are heavier than the $Z$ boson.

In this case the introduction of the $\phi_{X}$ scalar provides the
right elements to make the model safe. Moreover, concerning the neutrino
mass generation, from the Yukawa terms in Eq.~(\ref{lyukawa}) we
are able to construct the most general neutrino mass matrix, since
now all VEVs are different from zero.

\subsection{A $Z_{2}$ symmetry and dark matter}

Now let us consider the $Z_{2}$ symmetry again, after the introduction
of the scalar $\phi_{X}$. We had the field symmetry transformation
rules
\begin{equation}
Z_{2}(n_{R3})=-n_{R3},\quad Z_{2}(\Phi_{2})=-\Phi_{2},\quad Z_{2}(\phi_{3})=-\phi_{3},\label{z2}
\end{equation}
and all the other being even. It is easy to see that all the Hermitian
terms in the scalar potential involving $\phi_{X}$ are invariant
under $Z_{2}$. However, the non-Hermitian terms
\begin{equation}
-i\kappa_{H2X}(\Phi_{2}^{T}\tau_{2}H)(\phi_{X}^{\ast})^{2},\quad\beta_{X}(\phi_{X}^{\ast}\phi_{1})(\phi_{2}\phi_{3}),\quad\text{and}\quad\beta_{3X}(\phi_{X}^{\ast}\phi_{3}^{3}),\label{z2x}
\end{equation}
in $V_{B-L}^{X}$, are not invariant. We could change the $\phi_{X}$
transformation rule to odd, in order to have some of them invariant.
In this case, however, if we want to keep the Lagrangian invariant
under $Z_{2}$ after the SSB, we must have $V_{\phi_{X}}=0$, and
this is not an option since we need $V_{\phi_{X}}\neq0$ to have a
consistent model, as discussed above. Motivated by the possibility
of having DM candidates we impose the $Z_{2}$ symmetry to the entire
Lagrangian. Then, the terms in Eq.~(\ref{z2x}) will be removed from
the scalar potential and the only non-Hermitian terms allowed are
\begin{equation}
V_{B-L}^{NH}=\Phi_{1}^{\dagger}\Phi_{2}(\beta_{13}\phi_{1}\phi_{3}^{\ast}+\beta_{23}\phi_{2}^{\ast}\phi_{3})+\beta_{123}\phi_{1}\phi_{2}(\phi_{3}^{\ast})^{2}-i\kappa_{H1X}\Phi_{1}^{T}\tau_{2}H\phi_{X}+\textrm{H.c.}\label{potentialxnh}
\end{equation}
After the SSB, the $Z_{2}$ symmetry is not broken if we have $V_{\Phi_{2}}=V_{\phi_{3}}=0$,
and we have mass eigenstates that are also eigenstates of this symmetry.
However, we know from our previous analysis, before introducing $\phi_{X}$,
that this vacuum configuration challenges the safety of the model
due to a physical Goldstone and a light real scalar. Now, after introducing
$\phi_{X}$ we have four massless states in the neutral imaginary
sector. However, in this case, both of the physical massless states
are mainly singlets: $G_{F_{2}}^{0}$ is given by the same expression
in Eq.~(\ref{goldstones_fisicos2}), and
\begin{equation}
G_{F_{1}}^{0}\approx\frac{1}{\sqrt{7093}}\left(12\,\text{Im}\phi_{1}-15\,\text{Im}\phi_{2}+82\,\text{Im}\phi_{X}\right).
\end{equation}
In fact, for the parameter set we used above, the $G_{F_{1}}^{0}$
component in $\text{Im}H^{0}$ is $\approx2\times10^{-12}$ which
implies $g_{eeJ}\approx4\times10^{-18}$; thus it is safe with respect
to the star energy loss constraint. This main feature is due to the
introduction of the trilinear term $-i\kappa_{H1X}\Phi_{1}^{T}\tau_{2}H\phi_{X}$.
Qualitative arguments to explain this behavior can be given. The number
of $U(1)$ symmetries is the same in both situations, with and without
$\phi_{X}$, 4. Without $\phi_{X}$, we have two independent $U(1)$
symmetries involving only the doublets, say, $U(1)_{\sigma}$ acting
on $H$, and $U(1)_{\alpha}$ acting on $\Phi_{1,2}$. With $\phi_{X}$,
the trilinear term relates the $U(1)$ charge of $H$ to that of $\Phi_{1}$,
reducing the number of $U(1)$ symmetries involving only doublets
to just one, say, $U(1)_{\alpha}$ acting on $H$ and $\Phi_{1,2}$,
and at the same time, it introduces a new $U(1)$ symmetry acting
on $H$ and $\phi_{X}$, say, $U(1)_{\gamma}$. The number of broken
generators is the same in both situations, since we have the same
number of massless states; however, the origin of these physical massless
states is different. The introduction of $\phi_{X}$ works in a very
similar way to the singlet introduced in Refs.~\cite{axion1} and
\cite{axion2}\ to form the terms $H_{u}^{T}\tau_{2}H_{d}\phi$ and
$H_{u}^{T}\tau_{2}H_{d}\phi^{2}$, respectively, in order to make
the axion invisible.

The numerical spectra for the different scalar sectors, in GeV, are
$m_{C_{j}}=(1489.9,\,1330.3,\,0)$ for the charged scalar sector;
$m_{I_{i}}=(1479.7,\,1433.9,\,1318.9,\,0,\,0,\,0,\,0)$ for the neutral
imaginary sector, and $m_{R_{i}}=(1483.5,\,1479.7,\,1378.4,\,1378.4,\,1318.9,\,675.3,\,152.9)$
for the real scalar sector. The point here is that all real scalar
fields are now heavier than the $Z$ boson, avoiding in this way the
$Z$ invisible decay width constraint. Therefore, the model is safe
from these most severe constraints.

As the $Z_{2}$ symmetry still holds after the SSB, due to this particular
vacuum configuration, the model can present some DM candidates. In
general, a candidate must be the lightest particle odd under $Z_{2}$,
in order to be stable. In our case, it can be the lightest odd mass
eigenstate of the $n_{R3}$ or the lightest odd imaginary mass eigenstate,
or its odd real counterpart. This subject will be considered elsewhere.
Here we only estimate the relic abundance and the direct detection
cross section for the case of the fermionic cold DM candidate $n_{R3}$
(referred as $\chi$ from now on).

The most relevant annihilation process of $\chi$ occurs via the $t$-channel
exchange of $\Phi_{2}^{\pm}$($\Phi_{2}^{0}$) to charged (neutral)
leptons' final states. The thermally averaged $\chi$ annihilation
cross section, $\left\langle \sigma v\right\rangle $, is given by~\cite{bertone2005}
\begin{equation}
\left\langle \sigma v\right\rangle \approx a+b\,\left\langle v^{2}\right\rangle \approx\frac{1}{16\pi}\sum_{ij}G_{\textrm{eff},ij}^{2}\, c_{c}\, M_{\chi}^{2}\left\langle v^{2}\right\rangle .
\end{equation}
where $i,j=e,\mu,\tau$ and $c_{c}$ are the color factors, equal
to 1 for leptons. Also we have neglected the lepton masses. The $G_{\textrm{eff},ij}=D_{i3}D_{j3}^{*}/\left(m_{C_{1},R_{2}}^{2}+M_{\chi}^{2}\right)$
are the effective couplings, where $m_{C_{1}}\approx1489$ GeV ($m_{R_{2}}\approx1489$
GeV), the mass to be considered when charged (neutral) leptons are
produced. The relic abundance of $\chi$ is approximately given by~\cite{kolb1990}
\begin{equation}
\Omega h^{2}\approx\frac{1.04\times10^{-9}x_{f}}{M_{Pl}\sqrt{g_{*}}(a+3b/x_{f})},
\end{equation}
where, in this model, $g_{*}=107.75$ is the number of relativistic
degrees of freedom available at the freeze-out temperature, $T_{f}$,
and $x_{f}=M_{\chi}/T_{f}$ is given by
\begin{equation}
x_{f}=\ln\left[c\sqrt{\frac{45}{8}}\frac{g_{\chi}M_{\textrm{Pl}}M_{\chi}(a+6b/x_{f})}{2\pi^{3}\sqrt{x_{f}\, g_{*}}}\right],
\end{equation}
with $c=5/4$ and $g_{\chi}=2$. Using the following set of parameters,
$D_{e3}=0.06,\, D_{\mu3}=0.9,\, D_{e3}=1$, and for $M_{\chi}=750$
GeV, we find $x_{f}=24.81$ and $\Omega h^{2}=0.11$, which is in
agreement with the experimental bounds~\cite{komatsu2011}. The same
interaction allowing the $\chi$ annihilation in charged leptons,
which are proportional $D_{i3}$, also induces lepton flavor violation
(LFV) such as $\mu\rightarrow e\gamma$ and $\tau\rightarrow\mu\gamma$
(see below).

The next task is to compute the direct detection cross section. In
our case, the elastic scattering of $\chi$ with nuclei occurs via
the $t$-channel $\chi+N\rightarrow\,\chi+N$ process due to the exchange
of the scalar mass eigenstate $R_{7}$, which is the Higgs scalar
boson in the model with mass $m_{R_{7}}\approx152.9$ GeV. The spin-independent
cross section is given by~\cite{beltran2009}
\begin{equation}
\sigma_{\chi N}=\frac{4}{\pi}\frac{M_{\chi}^{2}m_{N}^{2}}{\left(M_{\chi}+m_{N}\right)^{2}}\left[Zf_{p}+(A-Z)f_{n}\right]^{2},
\end{equation}
where the effective couplings to protons and neutrons, $f_{p,n}$,
are
\begin{equation}
f_{p,n}=\sum_{q=u,d,s}\frac{G_{\textrm{eff},q}}{\sqrt{2}}f_{Tq}^{(p,n)}\frac{m_{p,n}}{m_{q}}+\frac{2}{27}f_{TG}^{(p,n)}\sum_{q=c,b,t}\frac{G_{\textrm{eff},q}}{\sqrt{2}}\frac{m_{p,n}}{m_{q}}.
\end{equation}
In this case $G_{\textrm{eff},q}=G_{0}\times m_{q}\equiv\left[C_{\phi_{2}R_{7}}C_{HR_{7}}M_{\chi}/\left(V_{\phi_{2}}V_{H}m_{R_{7}}^{2}\right)\right]\times m_{q}$,
where $C_{\phi_{2}R_{7}}\approx0.01$ and $C_{HR_{7}}\approx0.99$
are the coefficients relating the symmetry eigenstates ($\phi_{2}$,
$H$) to the relevant mass eigenstate $R_{7}$, respectively. By using
$f_{Tq}^{(p,n)}$ and $f_{TG}^{(p,n)}$ given in Ref.~\cite{ellis2000}
we find
\begin{equation}
\sigma_{\chi,p}\approx3\times10^{-7}\,\textrm{pb}\times\left[\frac{G_{\textrm{eff},q}\times(1\,\textrm{GeV}/\, m_{q})}{10^{-7}\,\textrm{GeV}^{-2}}\right]^{2},
\end{equation}
which gives $\sigma_{\chi,p}\approx4.74\times10^{-11}$ pb, for $M_{\chi}=750$
GeV, which is in agreement with the most recent present bounds~\cite{cdms2,xenon2010,xenon2011}.
The parameter set we have used in all the cases above is compatible
with the following requirements: (i) the constraint equations are
satisfied, (ii) all obtained masses are $m^{2}>0$, and (iii) results
for the already known fields are consistent with those of the SM at
the tree level.

Notice that, although we are considering a multi-Higgs model, the
values we have found for the lightest real scalar, the Higgs boson,
are in agreement with the last combined CDF and D0 results for the
ESM Higgs boson, which have excluded, at the 95\% C.L., a region at
high mass in $158<m_{H}<175$ GeV~\cite{cdfd0}.

\section{Neutrino masses}

The model without the $Z_{2}$ symmetry already has a satisfactory
solution to the neutrino masses, since we are able to construct a
general neutrino mass matrix. However, we are going to consider the
case with this symmetry because the model becomes more attractive
due to the presence of stable candidates to DM.

The Yukawa Lagrangian in Eq.~(\ref{lyukawa}) gives the following
neutrino mass terms:
\begin{equation}
-{\mathcal{L}}_{m_{\nu}}=\mathcal{D}_{im}\overline{\nu_{Li}}n_{Rm}V_{\Phi_{1}}+\mathcal{M}_{mn}\overline{(n_{m}^{c})_{L}}n_{Rn}V_{\phi_{1}}+\mathcal{M}_{33}\overline{(n_{3}^{c})_{L}}n_{R3}V_{\phi_{2}}+\textrm{H.c.},\label{nuterms}
\end{equation}
where $i,j=1,2,3$ (or $e,\,\mu,\,\tau$, respectively, when convenient)
and $m,n=1,2$. In matrix form Eq.~(\ref{nuterms}) reads

\begin{equation}
-{\mathcal{L}}_{m_{\nu}}=\left[\begin{array}{cc}
\overline{\nu_{L}} & \overline{(n^{c})_{L}}\end{array}\right]\left[\begin{array}{cc}
0 & M_{D}\\
M_{D}^{T} & M_{M}
\end{array}\right]\left[\begin{array}{c}
(\nu^{c})_{R}\\
n_{R}
\end{array}\right]\text{,}\label{matriz de massa}
\end{equation}
with
\begin{equation}
\nu_{L}=\left[\nu_{e}\,\,\nu_{\mu}\,\,\nu_{\tau}\right]_{L}^{T},\qquad n_{R}=\left[n_{1}\,\, n_{2}\,\, n_{3}\right]_{R}^{T}\text{.}
\end{equation}
The Majorana mass matrix ($M_{M}$) and the Dirac mass matrix ($M_{D}$)
are given by
\begin{equation}
M_{M}=V_{\phi_{1}}\left(\begin{array}{ccc}
\mathcal{M}_{11} & \mathcal{M}_{12} & 0\\
\mathcal{M}_{12} & \mathcal{M}_{22} & 0\\
0 & 0 & \frac{V_{\phi_{2}}}{V_{\phi_{1}}}\mathcal{M}_{33}
\end{array}\right),\qquad M_{D}=V_{\Phi_{1}}\left(\begin{array}{ccc}
\mathcal{D}_{11} & \mathcal{D}_{12} & 0\\
\mathcal{D}_{21} & \mathcal{D}_{22} & 0\\
\mathcal{D}_{31} & \mathcal{D}_{32} & 0
\end{array}\right),
\end{equation}
since $V_{\Phi_{2}}=V_{\phi_{3}}=0$, where $M_{M}=M_{M}^{T}$. For
$V_{\Phi_{1}}\ll V_{\phi_{1}}$, the mass matrix in Eq.~(\ref{matriz de massa})
can be diagonalized by an approximate scheme. The masses of the heavy
neutrinos are related to the energy scale of the VEVs of the singlets
$\phi_{1}$ and $\phi_{2}$, and are given by the eigenvalues of $M_{M}$:
\[
M_{1,2}=\frac{\left(\mathcal{M}_{11}+\mathcal{M}_{22}\right)\mp\sqrt{4\mathcal{M}_{12}^{2}+(\mathcal{M}_{11}-\mathcal{M}_{22})^{2}}}{2}V_{\phi_{1}},\qquad M_{3}=\mathcal{M}_{33}V_{\phi_{2}}.
\]
The masses of the light neutrinos are given by the eigenvalues of
the matrix
\begin{equation}
M_{\nu}\approx M_{D}M_{M}^{-1}M_{D}^{T},\label{lightnumass}
\end{equation}
which are
\begin{equation}
m_{1,2}=\frac{1}{2D_{M}}\left[\Delta\mp\sqrt{\Delta^{2}+r}\right]\frac{V_{\Phi_{1}}^{2}}{V_{\phi_{1}}},\qquad m_{3}=0,
\end{equation}
where the following definitions have been used:
\begin{eqnarray}
\overrightarrow{C}_{1} & = & (\mathcal{D}_{11},\mathcal{D}_{21},\mathcal{D}_{31}),\quad\overrightarrow{C}_{2}=(\mathcal{D}_{12},\mathcal{D}_{22},\mathcal{D}_{32}),\quad r=4D_{M}\left[\left(\mathcal{D}_{12}\overrightarrow{C}_{1}-\mathcal{D}_{11}\overrightarrow{C}_{2}\right)^{2}+D_{D}^{2}\right],\notag\\
\Delta & = & \mathcal{M}_{11}(\overrightarrow{C}_{2})^{2}+\mathcal{M}_{22}(\overrightarrow{C}_{1})^{2}-2\mathcal{M}_{12}(\overrightarrow{C}_{1}.\overrightarrow{C}_{2}),\quad D_{M}=\mathcal{M}_{12}^{2}-\mathcal{M}_{11}\mathcal{M}_{22},\\
D_{D} & = & \mathcal{D}_{21}\mathcal{D}_{32}-\mathcal{D}_{22}\mathcal{D}_{31}.\notag
\end{eqnarray}
The parameters in $M_{M}$ and $M_{D}$ have to be chosen in order
to have light neutrino masses consistent with the solar and atmospheric
experimental data. However, since there is no a standard procedure
to do that, we will present a particular solution to show that this
model can generate realistic active neutrino masses.

\subsection*{A particular solution}

From the experimental neutrino data it is found that the neutrino
mixing matrix is compatible with the so-called tribimaximal (TB) one~\cite{altarelli},
which is given by
\begin{equation}
U_{\textrm{TB}}=\left(\begin{array}{ccc}
\sqrt{\frac{2}{3}} & \frac{1}{\sqrt{3}} & 0\\
-\frac{1}{\sqrt{6}} & \frac{1}{\sqrt{3}} & -\frac{1}{\sqrt{2}}\\
-\frac{1}{\sqrt{6}} & \frac{1}{\sqrt{3}} & \frac{1}{\sqrt{2}}
\end{array}\right),\label{TB}
\end{equation}
and where it is assumed that the neutrino mixing angles are in a good
approximation given by $\sin^{2}\theta_{12}=1/3,\,\sin^{2}\theta_{23}=1/2$,
and $\sin^{2}\theta_{13}=0$. Working in a basis where the charged
lepton mass matrix is already diagonal, the $U_{\textrm{TB}}$ matrix
diagonalizes the light neutrino mass matrix in Eq.~(\ref{lightnumass}):
$U_{\textrm{TB}}^{T}M_{\nu}U_{\textrm{TB}}=\hat{M}_{\nu}=\text{diag}(m_{1},m_{2},m_{3})$.
It can be shown that the most general neutrino mass matrix that can
be exactly diagonalized by $U_{\textrm{TB}}$ has the form
\begin{equation}
M_{\textrm{TB}}=\left(\begin{array}{ccc}
x & y & y\\
y & x+\nu & y-\nu\\
y & y-\nu & x+\nu
\end{array}\right),\label{MTB}
\end{equation}
using the same notation as in Ref.~\cite{altarelli}.

The $M_{\textrm{TB}}$ mass eigenstates are
\begin{equation}
m_{1}=x-y,\qquad m_{2}=x+2y,\qquad m_{3}=2\nu+x-y.\label{mtbvals}
\end{equation}

The square mass differences $\Delta m_{\textrm{sol}}^{2}$ and $\Delta m_{\textrm{atm}}^{2}$,
needed to explain de solar and atmospheric neutrino anomalies, can
be obtained by imposing conditions on $x,y$, and $\nu$. The simplest
way to apply this analysis to our particular case is as follows. We
consider
\begin{equation}
\mathcal{M}_{11}=\mathcal{M}_{22}=\frac{V_{\phi2}}{V_{\phi_{1}}}\mathcal{M}_{33},\qquad\mathcal{M}_{12}=0,
\end{equation}
so that the Majorana and Dirac mass matrices are now given by
\begin{equation}
M_{M}=\mathcal{M}_{11}V_{\phi_{1}}\mathbf{1}_{3\times3};\qquad M_{D}=V_{\Phi_{1}}\left(\begin{array}{ccc}
\mathcal{D}_{11} & \mathcal{D}_{12} & 0\\
\mathcal{D}_{21} & \mathcal{D}_{22} & 0\\
\mathcal{D}_{31} & \mathcal{D}_{32} & 0
\end{array}\right).
\end{equation}
Then, the light neutrino mass matrix becomes
\begin{eqnarray}
M_{\nu} & = & M_{D}M_{M}^{-1}M_{D}^{T}=\frac{V_{\Phi_{1}}^{2}}{V_{\phi_{1}}}\frac{1}{\mathcal{M}_{11}}M_{D}M_{D}^{T}\notag\\
 & = & \frac{V_{\Phi_{1}}^{2}}{\mathcal{M}_{11}V_{\phi_{1}}}\left(\begin{array}{ccc}
\mathcal{D}_{11}^{2}+\mathcal{D}_{12}^{2} & \mathcal{D}_{11}\mathcal{D}_{21}+\mathcal{D}_{12}\mathcal{D}_{22} & \mathcal{D}_{11}\mathcal{D}_{31}+\mathcal{D}_{12}\mathcal{D}_{32}\\
\mathcal{D}_{11}\mathcal{D}_{21}+\mathcal{D}_{12}\mathcal{D}_{22} & \mathcal{D}_{21}^{2}+\mathcal{D}_{22}^{2} & \mathcal{D}_{21}\mathcal{D}_{31}+\mathcal{D}_{22}\mathcal{D}_{32}\\
\mathcal{D}_{11}\mathcal{D}_{31}+\mathcal{D}_{12}\mathcal{D}_{32} & \mathcal{D}_{21}\mathcal{D}_{31}+\mathcal{D}_{22}\mathcal{D}_{32} & \mathcal{D}_{31}^{2}+\mathcal{D}_{32}^{2}
\end{array}\right).\label{lightneutrinos}
\end{eqnarray}
The matrix above has a null determinant and, therefore, a zero mass
eigenstate. Hence, in order to make both matrices compatible, we must
have a vanishing eigenvalue in Eq.~(\ref{mtbvals}). We choose $m_{3}=0$,
i.e., $2\nu+x-y=0$, and, hence, $x+\nu=y-\nu$.

Comparing Eq.~(\ref{MTB}) with Eq.~(\ref{lightneutrinos}) we have
the following equations:
\begin{equation}
\frac{x}{K}=\mathcal{D}_{11}^{2}+\mathcal{D}_{12}^{2}\text{,}\label{xeq}
\end{equation}
\begin{equation}
\frac{y}{K}=\mathcal{D}_{11}\mathcal{D}_{21}+\mathcal{D}_{12}\mathcal{D}_{22}=\mathcal{D}_{11}\mathcal{D}_{31}+\mathcal{D}_{12}\mathcal{D}_{32}\text{,}\label{yeq}
\end{equation}
\begin{equation}
\frac{x+\nu}{K}=\mathcal{D}_{21}^{2}+\mathcal{D}_{22}^{2}=\mathcal{D}_{31}^{2}+\mathcal{D}_{32}^{2}\text{,}
\end{equation}
\begin{eqnarray}
\frac{y-\nu}{K} & = & \mathcal{D}_{21}\mathcal{D}_{31}+\mathcal{D}_{22}\mathcal{D}_{32}\text{,}\label{eqparameters}
\end{eqnarray}
where we have defined the dimensional constant $K=\frac{V_{\Phi_{1}}^{2}}{\mathcal{M}_{11}V_{\phi_{1}}}$.
A solution for the above equations is
\begin{equation}
\mathcal{D}_{21}=\mathcal{D}_{31}\text{,}\quad\text{and}\quad\mathcal{D}_{22}=\mathcal{D}_{32}\text{.}\label{gsol}
\end{equation}
From the above equations we see that the condition to have $m_{3}=0$,
$x+\nu=y-\nu$, is automatically satisfied. We have the following
equations to fit the atmospheric and solar neutrino data,
\begin{equation}
m_{1}=x-y,\qquad m_{2}=x+2y,\qquad m_{3}=0,
\end{equation}
and, therefore,
\begin{eqnarray}
\Delta m_{\textrm{sol}}^{2} & = & m_{2}^{2}-m_{1}^{2}=3y(2x+y)>0,\label{solar}\\
|\Delta m_{\textrm{atm}}^{2}| & = & |m_{3}^{2}-m_{1}^{2}|=(x-y)^{2}.\label{atmosferico}
\end{eqnarray}
Assuming that $x-y>0$ we have to solve the equations
\begin{equation}
3y(2x+y)=7.67\times10^{-5}\,\,(\text{eV})^{2},\quad\text{and}\,\,\, x-y=(2.4\times10^{-3})^{1/2}\,\,\text{eV},
\end{equation}
which are satisfied by $x=0.049\,248\,7$ and $y=0.000\,258\,887$,
in $\text{eV}$. The corresponding mass eigenvalues are then given
by $m_{1}=0.048\,989\,8,\, m_{2}=0.049\,766\,5$, and $m_{3}=0$,
in $\text{eV}$, showing an inverse hierarchy pattern. We can now
solve Eqs.~(\ref{xeq}) and (\ref{yeq}) for the $\mathcal{D}_{ij}$
parameters. In order to do that we have to know the value of the dimensional
constant $K$. For $V_{\Phi_{1}}=1\,\,\text{MeV}$, $V_{\phi_{1}}=1\,\,\text{TeV}$,
and assuming $\mathcal{M}_{11}=1$, we have $K=1\,\,\text{eV}$. Choosing
the input values $\mathcal{D}_{22}=0.25$ and $\mathcal{D}_{21}=0.15$,
we find $\mathcal{D}_{11}=0.190\,751$, and $\mathcal{D}_{12}=-0.113\,415$.
Experiments on $0\nu\beta\beta$ can put bounds on $|m_{ee}|$, and
the strongest one is $|m_{ee}|\,<\,0.26\,(0.34)\,\,\text{eV at}\,\,68\%\,(90\%)\,\text{C.L.}$~\cite{moscou}.
This quantity is related to the mass eigenvalues through $|m_{ee}|\ =|c_{13}^{2}(m_{1}c_{12}^{2}e^{i\delta_{1}}+m_{2}s_{12}^{2}e^{i\delta_{2}})+m_{3}e^{2i\phi_{CP}}s_{13}^{2}|$.
In our case, with no CP violation nor phases in the leptonic mixing
matrix, we find $|m_{ee}|\approx0.05$ eV. Future experiments, however,
expect to improve sensitivity up to $\approx0.01$ eV~\cite{newexpmee}.

The procedure we have followed for finding a particular solution for
the light neutrino masses can also be realized by using, instead of
the matrices given in Eqs.~(\ref{TB}) and (\ref{MTB}), the ones
given in Ref.~\cite{ma}, provided we make, in the notation of this
reference, $c=-d/2$, and the identifications $\nu=d-(a+b),\, y=d,\, x=a+2b-d$.
It results $-a=x-y+2\nu=m_{3}$, and we take a=0.

The results showed above demonstrate that the model is fully compatible
with the experimental neutrino data, and that light neutrino masses
can be generated neither appealing for very large energy scales nor
imposing fine-tuning. Now, we have to verify if the set of parameters
we have used above is in agreement with the LFV constraints coming
from a process like $l_{i}\rightarrow l_{j}+\gamma$, where $i=2,3=\mu,\tau$
and $j=1,2=e,\mu$, respectively. This model has one loop contributions
to such a process since charged leptons couple to charged scalars
and right-handed heavy neutrinos. The branching ratio is estimated
as~\cite{ma2001}
\begin{equation}
B\left(l_{i}\rightarrow l_{j}+\gamma\right)=\frac{96\pi^{3}\alpha}{G_{F}^{2}m_{l_{i}}^{4}}\left(\left|f_{M1}\right|^{2}+\left|f_{E1}\right|^{2}\right),
\end{equation}
where $\alpha\simeq1/137$ and $G_{F}\simeq1.16\times10^{-5}$ GeV$^{-2}$
is the Fermi constant and
\begin{equation}
f_{M1}=f_{E1}=\sum_{k=1}^{3}\frac{{\cal D}_{ik}{\cal D}_{jk}}{4\left(4\pi\right)^{2}}\frac{m_{l_{i}}^{2}}{m_{\Phi}^{2}}F_{2}\left(\frac{m_{N_{k}}^{2}}{m_{\Phi}^{2}}\right),
\end{equation}
with $F_{2}\left(x\right)$ being
\begin{equation}
F_{2}\left(x\right)=\frac{1-6x+3x^{2}+2x^{3}-6x^{2}\ln x}{6\left(1-x\right)^{4}}.
\end{equation}
Using the parameters needed to fit the neutrino masses and the ones
to estimate $\Omega h^{2}\simeq0.11$ (${\cal D}_{e3}\simeq0.06$,
${\cal D}_{\mu3}\simeq0.9$, ${\cal D}_{\tau3}\simeq1$, $m_{n_{R3}}=750$
GeV, $m_{C_{1}}\simeq m_{\Phi_{1}^{\pm}}=1.33$ TeV, $m_{C_{2}}=m_{\Phi_{2}^{\pm}}=1.48$
TeV), we can give an estimate for the branching ratio $B\left(\mu\rightarrow e+\gamma\right)\simeq7.9\times10^{-12}$
and $B\left(\tau\rightarrow\mu+\gamma\right)\simeq2.5\times10^{-9}$.
These values are in agreement with the present upper bounds $B\left(\mu\rightarrow e+\gamma\right)<1.2\times10^{-11}$
and $B\left(\tau\rightarrow\mu+\gamma\right)\simeq6.8\times10^{-8}$~\cite{brooks1999,aubert2005}.

The ratio between the VEVs we have used for finding the neutrino mass
eigenvalues is $V_{\Phi_{1}}/V_{\phi_{1}}=10^{-6}$. This is of the
same order as the ratio $m_{e}/m_{\text{top}}=Y_{e}/Y_{\text{top}}\approx10^{-6}$,
and it is comparable with $m_{u}/m_{\text{top}}=Y_{u}/Y_{\text{top}}\approx10^{-5}$.
We have chosen those values for $V_{\Phi_{1}}\,\text{and}\, V_{\phi_{1}}$
in order to have light neutrino masses without resorting to very tiny
neutrino Yukawa coupling constants, or fine-tuning, and, at the same
time, to have the $Z^{\prime}$ vector boson not extremely heavy.
This is a kind of seesaw mechanism where the heaviest scale, $V_{\phi_{1}}$,
is constrained by the $Z^{\prime}$ vector boson, which should be
not too heavy in order to not decouple from the spectrum. The light
scale, $V_{\Phi_{1}}$, is then used to fix the absolute neutrino
mass scale through the ratio $V_{\Phi_{1}}^{2}/V_{\phi_{1}}$. In
this picture we are substituting a hierarchy in the VEVs, which would
have a possible explanation based on the dynamics of the fields, for
one in the Yukawa coupling constants, for which we cannot find any
natural explanation. This is basically the philosophy behind the work
in Refs.~\cite{zee,anavicente}.

As we have discussed above, the absolute neutrino mass scale depends
on the ratio $V_{\Phi_{1}}^{2}/V_{\phi}$, where $V_{\Phi_{1}}$ is
a tiny value. Although this value can be affected by radiative corrections,
it can be argued that, when the $Z_{2}$ symmetry is considered, setting
$V_{\Phi_{1}}$ to a tiny value, at the tree level, is natural because
if it were in fact taken to be zero this would increase the symmetry
of the entire Lagrangian ('t Hooft's principle of naturalness). This
can be seen considering the constraint equations with $V_{\Phi_{1}}\rightarrow0$.
It implies that $\kappa_{H1X}=0$, since $V_{H}$ and $V_{\phi_{X}}$
differ from zero. Then the term $-i\kappa_{H1X}\Phi_{1}^{T}\tau_{2}H\phi_{X}$
does not appear in the scalar potential, Eq.~(\ref{potentialxnh}),
and the entire $Z_{2}$ invariant Lagrangian is now invariant under
an additional global quantum symmetry, say, $U(1)_{\zeta}$. A possible
$\zeta$-charge assignment is $\zeta(\nu_{eL},e_{L},e_{R},\Phi_{1,2})=-1$,
$\zeta(u_{L},d_{L},u_{R},d_{R})=1/3$, and $\zeta(n_{(1,2,3)R},\phi_{1,2,3})=0$.
Thus, it is expected that the VEV hierarchy will remain stable when
radiative corrections are taken into account.

\section{Conclusions}

In this paper we have studied in detail the scalar and the neutrino
Yukawa sectors of an extension of the electroweak standard model which
has an extra $U(1)$ gauge factor, as described in Sec.~II. We have
analyzed the scalar spectra of the potential given in Eq.~(\ref{potential1})
and found that it is inconsistent with the experimental data coming
from the star energy loss and the $Z$ invisible decay width. We would
like to stress that this is a general result for this scalar potential.

We find that the more suitable solution to this problem is the addition
of a new $SU(2)$ scalar singlet, called $\phi_{X}$ in the text.
The new terms introduced by $\phi_{X}$ are able to remove all the
physical Goldstone bosons and, at the same time, to have all the real
mass eigenstates heavier than the $Z$ boson. This solution is particularly
interesting since, in this case, all VEVs can be different from zero,
which allows for the construction of a general neutrino mass matrix.

In order to have a still more attractive model we consider the possibility
of having DM candidates by including a $Z_{2}$ symmetry. Before the
SSB the only fields having odd transformation under $Z_{2}$ are $n_{R3}$,
$\Phi_{2},\,\text{and}\,\phi_{3}$. $Z_{2}$ will still be a symmetry
if the scalar fields $\Phi_{2}\,\text{and}\,\phi_{3}$ do not develop
VEVs. Hence, after the SSB we will have states which are mass and
$Z_{2}$ eigenstates simultaneously. It opens the possibility of having
DM fields since the lighter $Z_{2}$ odd eigenstate will be stable.
Moreover, we show in a preliminary study that the fermionic field
$n_{R3}$ is a viable cold DM candidate.

We consider in detail the neutrino mass generation in the framework
of the model with the $Z_{2}$ symmetry. In this case we found an
inverted hierarchy compatible with the solar and atmospheric neutrino
data and the tribimaximal mixing matrix. Two appealing features are
(i) the absolute scale of the neutrino masses is obtained by a seesaw
mechanism at $\mathcal{O}\left(\text{TeV}\right)$ energy scale, which
is the scale of the first symmetry breaking, and (ii) the observed
mass-squared differences are obtained without resorting to fine-tuning
the neutrino Yukawa couplings.

The model has also some phenomenological implications. One of them
is the existence of an extra neutral vector boson, $Z^{\prime}$,
which can be in principle detected at the LHC or International Linear
Collider. In fact, there are studies showing that the $Z^{\prime}$
of this particular model can be distinguished from that of other models
by comparing, for instance, the forward-backward asymmetry for the
process $p+p\rightarrow\mu^{+}+\mu^{-}+X$ as a function of the dilepton
invariant mass, or the muon transverse momentum distribution at the
LHC~\cite{coutinho2011}, and the same asymmetry for the process
$e^{+}+e^{-}\rightarrow f+\overline{f}$ ($f=q,\, l$) at International
Linear Collider~\cite{fortes2010}. At first glance, another interesting
feature is that the model seems to indicate that the LFV and DM are
closely related. It implies that when the parameters are appropriate
to satisfy the DM requirements, the LFV is relatively close to the
present experimental bounds. In this way, the model can be confronted
by the next generation of LFV experiments.

\acknowledgements B. L. Sánchez--Vega was supported by CAPES. We
are grateful to E. Pontón and V. Pleitez for valuable discussions.

\appendix

\section{The constraint equations}

\label{sec:constraint}Here we show the constraint equations for the
scalar potential given in Eq.~(\ref{potential1}) plus the terms
after the $\phi_{X}$ introduction and without the $Z_{2}$ symmetry.
These equations are obtained by considering, after the spontaneous
symmetry breaking, the linear terms ($t_{\varphi}\varphi$) in the
scalar potential, and the solutions to the equations $t_{\varphi}=0$
are the critical points of the scalar potential.
\begin{eqnarray*}
 &  & t_{H}=V_{H}\left(2\text{\ensuremath{\lambda_{H}}}V_{H}^{2}+\Lambda_{H1}V_{\Phi_{1}}^{2}+\Lambda_{H2}V_{\Phi_{2}}^{2}+\Lambda_{Hs1}V_{\phi_{1}}^{2}+\Lambda_{Hs2}V_{\phi_{2}}^{2}+\Lambda_{Hs3}V_{\phi_{3}}^{2}+\Lambda_{HsX}V_{\phi_{X}}^{2}\right.\\
 &  & \left.-2\text{\ensuremath{\mu_{H}^{2}}}\right)-\sqrt{2}\kappa_{H1X}V_{\Phi_{1}}V_{\phi_{X}}-\kappa_{H2X}V_{\Phi_{2}}V_{\phi_{X}}^{2},
\end{eqnarray*}
\begin{eqnarray*}
 & t_{\Phi_{1}}= & V_{\Phi_{1}}\left(\Lambda_{H1}V_{H}^{2}+2\text{\ensuremath{\lambda_{11}}}V_{\Phi_{1}}^{2}+(\text{\ensuremath{\lambda_{12}}}+\text{\ensuremath{\lambda_{12}^{\prime}}})V_{\Phi_{2}}^{2}+\Lambda_{11}^{\prime}V_{\phi_{1}}^{2}+\Lambda_{12}^{\prime}V_{\phi_{2}}^{2}+\Lambda_{13}^{\prime}V_{\phi_{3}}^{2}+\Lambda_{1X}^{\prime}V_{\phi_{X}}^{2}\right.\\
 &  & \left.-2\text{\ensuremath{\mu_{11}^{2}}}\right)-\sqrt{2}\text{\ensuremath{\kappa_{H1X}}}V_{H}V_{\phi_{X}}+V_{\Phi_{2}}V_{\phi_{3}}(\beta_{13}V_{\phi_{1}}+\beta_{23}V_{\phi_{2}}),
\end{eqnarray*}
\begin{eqnarray*}
 & t_{\Phi_{2}}= & V_{\Phi_{2}}\left(\Lambda_{H2}V_{H}^{2}+(\lambda_{12}+\lambda_{12}^{\prime})V_{\Phi_{1}}^{2}+2\text{\ensuremath{\lambda_{22}}}V_{\Phi_{2}}^{2}+\Lambda_{21}^{\prime}V_{\phi_{1}}^{2}+\Lambda_{22}^{\prime}V_{\phi_{2}}^{2}+\Lambda_{23}^{\prime}V_{\phi_{3}}^{2}+\Lambda_{2X}^{\prime}V_{\phi_{X}}^{2}\right.\\
 &  & \left.-2\text{\ensuremath{\mu_{22}^{2}}}\right)-\text{\ensuremath{\kappa_{H2X}}}V_{H}V_{\phi_{X}}^{2}+V_{\Phi_{1}}V_{\phi_{3}}(\beta_{13}V_{\phi_{1}}+\beta_{23}V_{\phi_{2}}),
\end{eqnarray*}
\begin{eqnarray*}
 & t_{\phi_{1}}= & V_{\phi_{1}}\left(\Lambda_{Hs1}V_{H}^{2}+\Lambda_{11}^{\prime}V_{\Phi_{1}}^{2}+\Lambda_{21}^{\prime}V_{\Phi_{2}}^{2}+2\text{\ensuremath{\lambda_{s1}}}V_{\phi_{1}}^{2}+\Delta_{12}V_{\phi_{2}}^{2}+\Delta_{13}V_{\phi_{3}}^{2}+\Delta_{1X}V_{\phi_{X}}^{2}-2\text{\ensuremath{\mu_{s1}^{2}}}\right)\\
 &  & +\beta_{13}V_{\Phi_{1}}V_{\Phi_{2}}V_{\phi_{3}}+V_{\phi_{2}}V_{\phi_{3}}(\beta_{123}V_{\phi_{3}}+\beta_{X}V_{\phi_{X}}),
\end{eqnarray*}
\begin{eqnarray*}
 & t_{\phi_{2}}= & V_{\phi_{2}}\left(\Lambda_{Hs2}V_{H}^{2}+\Lambda_{12}^{\prime}V_{\Phi_{1}}^{2}+\Lambda_{22}^{\prime}V_{\Phi_{2}}^{2}+\Delta_{12}V_{\phi_{1}}^{2}+2\text{\ensuremath{\lambda_{s2}}}V_{\phi_{2}}^{2}+\Delta_{23}V_{\phi_{3}}^{2}+\Delta_{2X}V_{\phi_{X}}^{2}-2\text{\ensuremath{\mu_{s2}^{2}}}\right)\\
 &  & +\beta_{23}V_{\Phi_{1}}V_{\Phi_{2}}V_{\phi_{3}}+V_{\phi_{1}}V_{\phi_{3}}(\beta_{123}V_{\phi_{3}}+\beta_{X}V_{\phi_{X}}),
\end{eqnarray*}
\begin{eqnarray*}
 & t_{\phi_{3}}= & V_{\phi_{3}}\left(\Lambda_{Hs3}V_{H}^{2}+\Lambda_{13}^{\prime}V_{\Phi_{1}}^{2}+\Lambda_{23}^{\prime}V_{\Phi_{2}}^{2}+\Delta_{13}V_{\phi_{1}}^{2}+\Delta_{23}V_{\phi_{2}}^{2}+2\text{\ensuremath{\lambda_{s3}}}V_{\phi_{3}}^{2}+\Delta_{3X}V_{\phi_{X}}^{2}\right.\\
 &  & \left.+3\beta_{3X}V_{\phi_{3}}V_{\phi_{X}}-2\text{\ensuremath{\mu_{s3}^{2}}}\right)+V_{\Phi_{1}}V_{\Phi_{2}}(\beta_{13}V_{\phi_{1}}+\beta_{23}V_{\phi_{2}})+V_{\phi_{1}}V_{\phi_{2}}(2\beta_{123}V_{\phi_{3}}+\beta_{X}V_{\phi_{X}}),
\end{eqnarray*}
\begin{eqnarray*}
 & t_{\phi_{X}}= & V_{\phi_{X}}\left(\Lambda_{HsX}V_{H}^{2}+\Lambda_{1X}^{\prime}V_{\Phi_{1}}^{2}+\Lambda_{2X}^{\prime}V_{\Phi_{2}}^{2}+\Delta_{1X}V_{\phi_{1}}^{2}+\Delta_{2X}V_{\phi_{2}}^{2}+\Delta_{3X}V_{\phi_{3}}^{2}+2\text{\ensuremath{\lambda_{sX}}}V_{\phi_{X}}^{2}\right.\\
 &  & \left.-2\text{\ensuremath{\kappa_{H2X}}}V_{H}V_{\Phi_{2}}-2\text{\ensuremath{\mu_{sX}^{2}}}\right)-\sqrt{2}\text{\ensuremath{\kappa_{H1X}}}V_{H}V_{\Phi_{1}}+\beta_{X}V_{\phi_{1}}V_{\phi_{2}}V_{\phi_{3}}+\beta_{3X}V_{\phi_{3}}^{3}.
\end{eqnarray*}

\end{document}